\documentclass[amsmath,amssymb,superscriptaddress,nobalancelastpage,aps,prl,twocolumn]{revtex4-2}

\usepackage{graphicx}
\usepackage{varioref}
\usepackage{xr-hyper}
\usepackage{xcolor}
\usepackage{nicefrac}
\usepackage{xfrac}
\usepackage{hyperref}
\hypersetup{colorlinks,linkcolor=blue,urlcolor=blue,citecolor=blue}
\usepackage{ulem}
\usepackage{lineno}
\usepackage{amsmath} 
\usepackage{amssymb}
\usepackage{graphicx}
\usepackage{dcolumn}
\usepackage{bm}
\usepackage{color}
\definecolor{Green}{rgb}{0, 0.7, 0}

\usepackage{pifont}
\usepackage{natbib}
\usepackage{hyperref}
\hypersetup{
colorlinks = true,
urlcolor   = blue,
linkcolor  = blue,
citecolor  = blue
}

\begin{document}

\title{Anatomy of anomalous Hall effect due to magnetic fluctuations}

\author{Ola~Kenji~Forslund}
 \email{ola.forslund@physik.uzh.ch}
\affiliation{Physik-Institut, Universität Zürich, Winterthurerstrasse 190, CH-8057 Zürich, Switzerland}
\affiliation{Department of Physics and Astronomy, Uppsala University, Box 516, SE-75120 Uppsala, Sweden}

\author{Xiaoxiong~Liu}
\affiliation{Shenzhen Institute for Quantum Science and Engineering and Department of Physics, Southern University of Science and Technology (SUSTech), Shenzhen 518055, China}
\affiliation{Quantum Science Center of Guangdong-Hong Kong-Macao Greater Bay Area (Guangdong), Shenzhen 518045, China}
\affiliation{Shenzhen Key Laboratory of Quantum Science and Engineering, Shenzhen 518055, China}
\affiliation{International Quantum Academy, Shenzhen 518048, China}

\author{Soohyeon~Shin}
\affiliation{Laboratory for Multiscale Materials Experiments, PSI Center for Neutron and Muon Sciences, Forschungsstrasse 111, 5232 Villigen PSI, Switzerland}

\author{Chun~Lin}
\affiliation{Physik-Institut, Universität Zürich, Winterthurerstrasse 190, CH-8057 Zürich, Switzerland}

\author{Masafumi~Horio}
\affiliation{Physik-Institut, Universität Zürich, Winterthurerstrasse 190, CH-8057 Zürich, Switzerland}

\author{Qisi~Wang}
\affiliation{Physik-Institut, Universität Zürich, Winterthurerstrasse 190, CH-8057 Zürich, Switzerland}

\affiliation{Department of Physics, The Chinese University of Hong Kong, Shatin, Hong Kong, China}

\author{Kevin~Kramer}
\affiliation{Physik-Institut, Universität Zürich, Winterthurerstrasse 190, CH-8057 Zürich, Switzerland}

\author{Saumya~Mukherjee}
\affiliation{Diamond Light Source, Didcot OX11 0DE, United Kingdom}

\author{Timur~Kim}
\affiliation{Diamond Light Source, Didcot OX11 0DE, United Kingdom}

\author{Cephise~Cacho}
\affiliation{Diamond Light Source, Didcot OX11 0DE, United Kingdom}

\author{Chennan~Wang}
\affiliation{Paul Scherrer Institute, Laboratory for Muon Spin Spectroscopy, CH-5232 PSI Villigen, Switzerland}

\author{Tian~Shang}
\affiliation{Key Laboratory of Polar Materials and Devices (MOE), School of Physics and Electronic Science, East China Normal University, Shanghai 200241, China}

\author{Victor Ukleev}
\affiliation{Laboratory for Neutron Scattering and Imaging (LNS), PSI Center for Neutron and Muon Sciences, Forschungsstrasse 111, 5232 Villigen PSI, Switzerland}

\affiliation{Helmholtz-Zentrum Berlin für Materialien und Energie, Berlin, Germany}

\author{Jonathan~S.~White}
\affiliation{Laboratory for Neutron Scattering and Imaging (LNS), PSI Center for Neutron and Muon Sciences, Forschungsstrasse 111, 5232 Villigen PSI, Switzerland}

\author{Pascal~Puphal}
\affiliation{Laboratory for Multiscale Materials Experiments, PSI Center for Neutron and Muon Sciences, Forschungsstrasse 111, 5232 Villigen PSI, Switzerland}
\affiliation{Max-Planck-Institute for Solid State Research, Heisenbergstraße 1, 70569 Stuttgart, Germany}

\author{Yasmine~Sassa}
\affiliation{Department of Physics, Chalmers University of Technology, SE-412 96 G\"oteborg, Sweden}

\author{Ekaterina~Pomjakushina}
\affiliation{Laboratory for Multiscale Materials Experiments, PSI Center for Neutron and Muon Sciences, Forschungsstrasse 111, 5232 Villigen PSI, Switzerland}

\author{Titus~Neupert}
\affiliation{Physik-Institut, Universität Zürich, Winterthurerstrasse 190, CH-8057 Zürich, Switzerland}

\author{Johan~Chang}
\affiliation{Physik-Institut, Universität Zürich, Winterthurerstrasse 190, CH-8057 Zürich, Switzerland}

\date{\today}

\begin{abstract}
The anomalous Hall {\color{black} e}ffect (AHE) has emerged as a key indicator of time-reversal symmetry breaking (TRSB) and topological features in electronic band structures. Absent of a magnetic field, the AHE requires spontaneous TRSB but has proven hard to probe due to averaging over domains. The anomalous component of the Hall effect is thus frequently derived from extrapolating the magnetic field dependence of the Hall response. We show that discerning whether the AHE is an intrinsic property of the field free system becomes intricate in the presence of strong magnetic fluctuations. {\color{black}As a study case,} we use the Weyl semimetal PrAlGe, where TRSB can be toggled via a ferromagnetic transition, providing a transparent view of the AHE's topological origin.  Through a combination of thermodynamic, transport and muon spin relaxation measurements, we contrast the behaviour below the ferromagnetic transition temperature to that of strong magnetic fluctuations above. Our results {\color{black}on PrAlGe provide general insights into the} interpretation of anomalous Hall signals in systems where TRSB is debated, such as families of Kagome metals or certain transition metal dichalcogenides. 
\end{abstract}



\maketitle
Intrinsic anomalous Hall effect (AHE) \cite{Hasan2010, Bernevig2022}, characterized by transverse Hall conductance in the absence of an external magnetic field, has garnered significant attention as a probe of time-reversal symmetry breaking (TRSB) and topological features in electronic systems \cite{Machida2010, Nakatsuji2015, Son2019, Ho2021}. Unlike extrinsic AHE, which is driven by asymmetric scattering processes, intrinsic AHE (from here on AHE) is rooted in the non-zero integral of the Berry curvature \cite{Nagaosa2010}. {\color{black}Understanding the microscopic origins behind an observed Hall signal is vital, as spontaneous TRSB is a prerequisite for AHE. However, despite these profound implications,} detecting and analyzing AHE is challenging due to magnetic domain formation. Therefore, externally applied magnetic fields are commonly utilised to lift the domain degeneracy. The AHE is in this case derived from extrapolating observed Hall resistance to the zero-field limit~\cite{Nagaosa2010}. 

In ferromagnetic Weyl semimetals, the intrinsic anomalous Hall conductivity for small magnetic fields ($\boldsymbol{B}$) is proportional to $\boldsymbol{B}$, the sum of externally applied field ($\boldsymbol{H}$) and magnetisation ($\boldsymbol{M}$) (other Weyl semimetals exist in which the average magnetisation is not the suitable TRSB order parameter to serve as a proxy for the AHE \cite{Nayak2016}). 
Therefore, $\boldsymbol{H}$ and temperature ($T$) dependence of $\boldsymbol{B}$ serve as a estimate for the expected anomalous Hall conductivity. This indirect approach can, however, obscure whether the observed AHE is intrinsic or induced by the applied magnetic field. Differentiating these two scenarios is difficult and hence complicating the interpretation of many experiments \cite{Liang2018, Yang2020, Mielke2022, Zheng2023, Reichlova2024, Wei2024}. 


{\color{black}The case of} PrAlGe, a ferromagnetic (FM) Weyl semimetal, offers a unique opportunity to disentangle these effects. PrAlGe is a well-characterized topological ferromagnet ($T_{\rm C} = 15$~K) stabilized in a non-centrosymmetric structure (space group $I4_1md$, \# 109). Due to broken inversion symmetry, Weyl points are expected to persist at all temperatures \cite{Chang2018}. By utilizing temperature as a control parameter, we can precisely navigate between the ferromagnetic state (with TRSB) and the paramagnetic state (without TRSB), allowing us to systematically explore the AHE's dependence on magnetic order,  effectively distinguishing between intrinsic and external contributions.

In this letter, we report a Hall resistance study across the ferromagnetic transition. Consistent with previous reports, an intrinsic Hall effect is observed in the ordered state, which is derived from zero-field extrapolation. We demonstrate that a similar extrapolated response is found in the paramagnetic phase. Muon spin relaxation ($\mu^+$SR) measurements reveal the absence of short-range magnetic order or domain formations, ruling out TRSB above the ferromagnetic ordering temperature. {\color{black} In the presence of short-range magnetic order or domain formations, $\mu^+$SR spectra typically exhibit either a sum of highly relaxing Gaussian-like functions or a combination of exponential terms. However, our $\mu^+$SR results do not show these features and the data is consistent with the presence of dynamic paramagnetic fluctuations.} Instead, we show that the application of a magnetic field in presence of strong ferromagnetic fluctuations can induce a finite magnetisation, as an immediate consequence of the proximate mean-field transition. This in turn produce a finite zero-field extrapolated Hall effect. We therefore demonstrate that an extrapolated AHE is not a definitive indicator of TRSB. As such, our study establishes a framework for accurately interpreting the anomalous Hall measurements.

Single-crystalline PrAlGe (based on Pr (99.9\%, pieces, ChemPur), Ge (99.999\%, pieces, Alfa Aesar), and Al (99.99\%, granules, ChemPur)) were grown using {\color{black} both} the Al-flux method and floating-zone methods as described in Refs.~\onlinecite{Paul2016, Puphal2019}. {\color{black} Flux grown single crystals were used for the electrical and magnetic susceptibility measurements, whereas the floating-zone crystals were used for angle resolved photo emission spectroscopy (ARPES) measurements. Polycrystalline samples for the $\mu^+$SR measurements were prepared by arc melting, as described in Ref.~\onlinecite{Puphal2020}. All samples were characterized through X-ray diffraction, resistivity, and DC magnetization, confirming their quality and consistent ferromagnetic transition temperature of $T_{\rm C}\simeq$ 15~K, ensuring comparable magnetic properties \cite{Destraz2020}. The experimental setup of each technique is described in end matter and summerised in table~\ref{table:table}. 
}

\begin{figure}[ht]
  \begin{center}
     \includegraphics[keepaspectratio=true,width=90 mm]{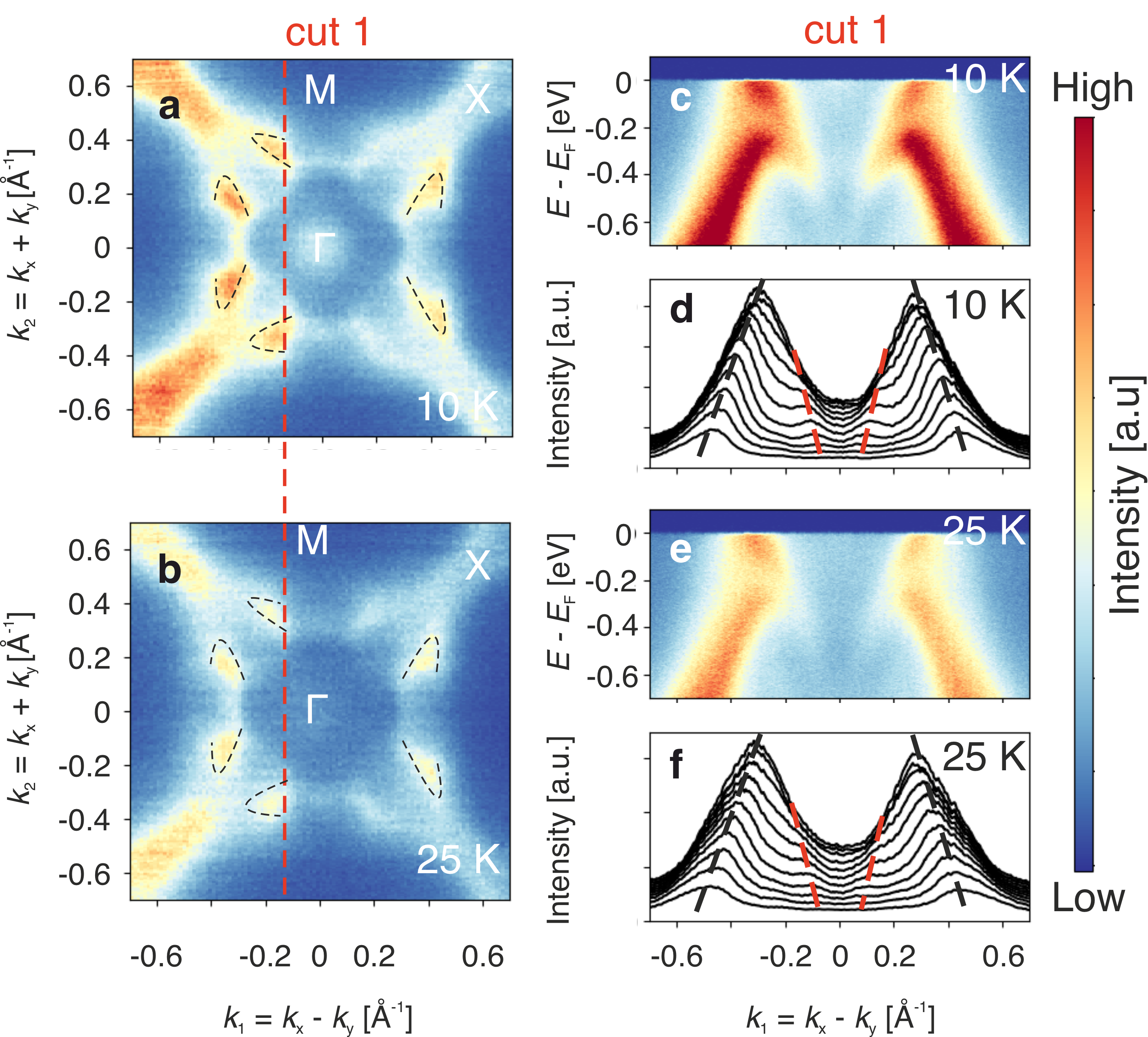}
 \end{center}
    \caption{(a, b) Angular resolved photo emission spectroscopy (ARPES) intensity recorded at the Fermi level with $h\nu=34$~eV photons for temperatures as indicated. The dashed black lines corresponds to the $U$-shaped Fermi arcs reported in Ref.~\onlinecite{Sanchez2020}. (c, e) Band dispersion through the $U$-shaped Fermi arc (cut 1) for $T=10$~K and 25~K, respectively. (d, f) Momentum distribution curves from (c, e) at selected binding energies. The dashed lines are guides to the eye. 
    }
    \label{fig:ARPES}
\end{figure}

To confirm the electronic band structure below and above $T_{\rm C}$, we carried out ARPES. Previous {\color{black}density functional theory} calculations \cite{Chang2018} predicted Weyl points near the zone center. A photon energy ($h\nu$) scan determines that $h\nu=34$ and 28~eV cuts through the zone center and boundary, respectively (see Fig.~\ref{fig:p_xy}(c)). Figure~\ref{fig:ARPES}(a, b) show the measured Fermi surface (FS) collected using $hv = 34$~eV photons for temperatures below and above $T_C$, as indicated. The observed FS is consistent with a previously reported study conducted at $T \sim 11$~K \cite{Sanchez2020}, that however didn't address the paramagnetic state ($T>T_{\rm C}$). $U$-shaped contours, present at the Fermi level, were interpreted as Fermi arcs. We find that the reported Fermi arcs are also present at $T = 25$~K (Fig.~\ref{fig:ARPES}(b)), which confirms the existence of Weyl points even in the paramagnetic phase.

A vertical momentum cut (cut 1: parallel to $\Gamma-M$) at $k_1\simeq -0.18$~\AA$^{-1}$ passes through a pair of “U” states and two bands approaching the Fermi level are observed (Fig.~\ref{fig:ARPES}(c, e)). Based on the predicted location of the Weyl nodes, one of these linear bands were previously interpreted as left and right moving two chiral edge modes \cite{Sanchez2020}. This is further enhanced in their respective momentum distribution curves (MDCs) for selected binding energies (Fig.~\ref{fig:ARPES}(d, f)). Given the lack of any significant temperature dependence in the {\color{black}band structure}, these Weyl points are most likely originating from the inversion symmetry breaking rather than TRSB from the FM order. {\color{black} The difference in intensity between the two temperatures is attributed to surface aging.} 

\begin{figure}[t]
  \begin{center}
     \includegraphics[keepaspectratio=true,width=90 mm]{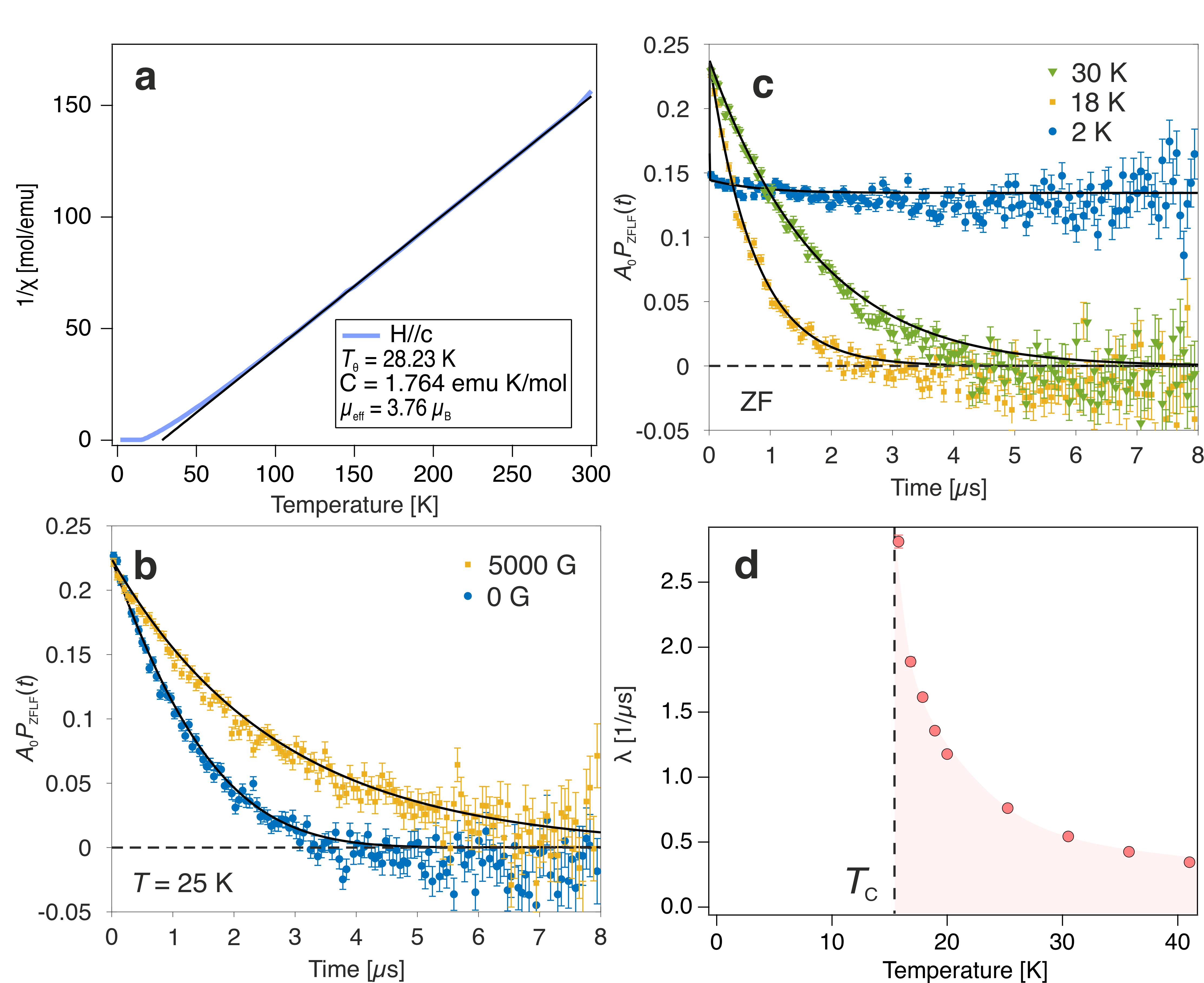}
 \end{center}
    \caption{(a) Inverse magnetic susceptibility ($\chi$) as a function of temperature. The solid line is a linear fit based on Curie-Weiss law. Fit parameters are shown in the figure inset. (b) Zero and longitudinal field (ZF and LF) muon time spectra collected at $T=25$~K. Solid lines are fits described in the main text {\color{black} and end matter appendix~D}. (c) Zero field muon time spectra for selected temperatures. The solid lines, above $T_{\rm C}$, are fits to extract the muon spin relaxation rate ($\lambda$) -- see main text and {\color{black}end matter appendix~D}. (d) Temperature dependence of $\lambda$ above $T_{\rm C}$. The coloured area is guide to the eye. The vertical dash lines indicate the critical temperature $T_{\rm C} = 15$~K.}
    
    \label{fig:muon}
\end{figure}

To characterise the magnetic properties of PrAlGe, temperature dependent inverse magnetic susceptibility ($\chi$) is shown in Fig.~\ref{fig:muon}(a). A Curie-Weiss (CW) behaviour is observed for $T > 80$~K and the positive value of the Weiss constant suggest FM interactions, consistent with previous reports \cite{Meng2019, Destraz2020}. Deviation from linearity is observed below 80~K, well above the FM ordering temperature $T_{\rm C}=15$~K. This FM state is stabilised via spin polarised $f$-electron states aligned along the {\color{black} FM easy-axis ($c$-axis) \cite{Destraz2020}. Deviations from the CW law may signal a gradual slowing down of dynamic spin-spin correlations due to thermal fluctuations. This involves a relatively large magnetic interaction strength, $J\sim k_{\rm B}T$, likely dominated by interactions along the $c$-axis \cite{Destraz2020}, where spins interact and fluctuate coherently over short timescales. This anisotropic spin interactions along the $c$-axis contribute to enhanced spin fluctuations across a broad temperature range \cite{Coey2010}. As the system approaches $T_{\rm C}$, these fluctuations become slower leading to the development of effective internal magnetic fields, which facilitates an induced magnetisation under an external magnetic field.}

To confirm the outlined scenario, we measured zero field (ZF) $\mu^+$SR time spectra as a function of temperature (Fig.~\ref{fig:muon}(c)). $\mu^+$SR is a sensitive magnetic probe for detecting both static and dynamic local magnetic fields in the MHz-GHz range. The ZF time spectra display an exponential depolarisation. Using the fit function $A_{0}P_{ZFLF}(t) = AG^{{\color{black}LF}SGKT}(\Delta,t, {\color{black}B_{\rm LF}})e^{-\lambda t}$, with {\color{black}$G^{LFSGKT}(\Delta,t,B_{\rm LF})$} being the {\color{black}longitudinal field} static Gaussian Kubo-Toyabe function \cite{Yaouanc2011}, we find a relaxation rate of $\lambda\simeq0.7~\mu$s$^{-1}$ {\color{black}(the fit function is further explained and justified in end matter appendix~D). In ZF, $B_{\rm LF}=0$ and $G^{LFSGKT}$ reduces into $G^{SGKT}$.} As $G^{SGKT}(\Delta,t)$ represents the depolarisation originating from nuclear magnetic moments (mostly $I_{Al}^{27}=5/2 $ and $I_{Pr}^{131}=5/2$), it is expected to be independent of temperature {\color{black} and a value of $\Delta=0.15$~$\mu$s$^{-1}$ was obtained (see end matter appendix~D)}. This implies that any temperature dependence originates from the $e^{-\lambda t}$ term. This term represents the depolarisation originating from {\color{black}electronic moments} and its temperature dependence is presented in Fig.~\ref{fig:muon}(d). The relaxation rate ($\lambda$) exhibits the maximum value close to $T_{\rm C}$, and gradually decreases with increasing temperature. Since the relaxation rate is inversely proportional to the spin-spin correlation frequency \cite{Yaouanc2011}, the increase close to $T_{\rm C}$ is understood as a critical slowing down of fluctuating {\color{black} electronic moments} close to $T_{\rm C}$, and is in-line with the described scenario based on the magnetisation measurements. This kind of increase close to $T_{\rm C}$ is commonly observed in many second order magnetic transitions \cite{Forslund2020_Na, Forslund2020, Higa2021}. 

The exponential relaxation in the time spectra (Fig.~\ref{fig:muon}(b,c)) suggest a dynamic state. However, even static field distribution may in certain cases yield exponential like relaxation if the internal field distribution is Lorentzian rather than Gaussian. Here, we confirm the dynamic origin via a longitudinal field (LF) measurement (Fig.~\ref{fig:muon}(b)). If the internal field is static, application of weak LF should decouple the spectra ($i.e.$ suppress the depolarisation so that effectively $\lambda\simeq 0$~$\mu$s$^{-1}$). However, even a strong LF = 5000~G does not decouple the spectra and the internal field is confirmed to be dynamic {\color{black} (see end matter appendix~D)}. This is in-line with temperature dependent resistivity study in which measurements across $T_{\rm C}$ exhibits a sharp drop \cite{Destraz2020}. This drop is consistent with suppressed scattering from magnetic fluctuations in the FM phase, fluctuations that was prevailing in the paramagnetic state. It is also noted that previous neutron \cite{Destraz2020} and $\mu^+$SR \cite{Puphal2020} measurements suggested the formation of nano-sized magnetic domain wall spin textures in the FM phase. We may however conclude from our data that similar magnetic domains or smaller patches of long range magnetic order can be excluded above $T_{\rm C}$. This statement is based on the fact that Gaussian like relaxation or oscillations are absent in the time spectra.

\begin{figure}[t]
  \begin{center}
     \includegraphics[keepaspectratio=true,width=90 mm]{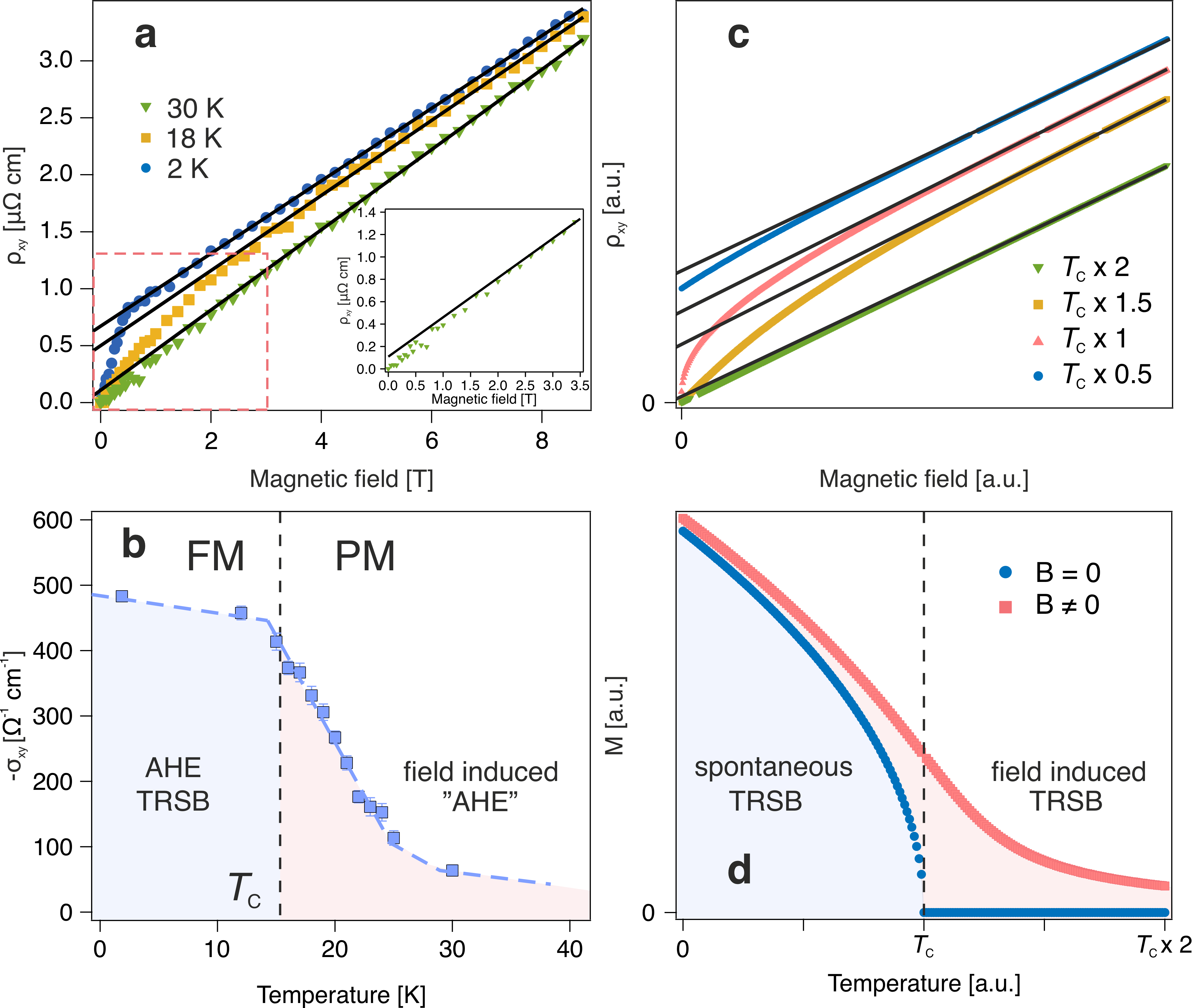}
 \end{center}
    \caption{(a) Hall resistivity ($\rho_{xy}$) as a function of magnetic field for temperatures as indicated. Solid lines are linear fits in the magnetic field range $H > 2.5$~T. The inset shows the low magnetic field range as indicated by the dashed red rectangle in the main figure. (b) Temperature dependence of $\sigma_{xy} (H\rightarrow0)$, obtained {\color{black}from} linear fits in (a) -- described in the main text. Below the critical temperature ($T_{\rm C}$), spontaneous time reversal symmetry breaking (TRSB) asserts the measured $\sigma_{xy}$ to be anomalous Hall conductivity. Above $T_{\rm C}$, the TRS is preserved and the measured $\sigma_{xy}$ is instead an artifact from the applied magnetic field. The coloured dashed lines are guide to the eye. (c) The calculated Hall resistivity curves as a function of magnetic field for selected temperature as indicated. (d) Calculated magnetisation as a function of temperature for zero and finite magnetic fields. The calculations are explained in the main text and supplementary materials {\color{blue} \cite{SM}.} }
    
    \label{fig:transport}
\end{figure}

Having established both macroscopically and microscopically the existence of spin fluctuations above $T_{\rm C}$, we next present temperature dependent study of the AHE. The temperature dependence is determined from transverse resistivity ($\rho_{xy}$) isotherms, where a linear behaviour is observed at higher fields (Fig.~\ref{fig:transport}(a); {\color{black}see Fig.~\ref{fig:p_xy}(a,b)} for more temperatures). This linear behaviour is present even for temperatures above $T_{\rm C}=15$~K. The Hall resistivity in FM materials contain contributions from the ordinary ($\rho^O_{xy}$) and anomalous Hall resistivity ($\rho^A_{xy}$), where the intercept at zero field of the linear behaviour extrapolated from higher fields defines $\rho^A_{xy}$ \cite{Nagaosa2010}. The transverse conductivity can be evaluated based on $\rho^A_{xy}$ via $\sigma_{xy} = -\rho^A_{xy}/(\rho_{xx}^2+\rho_{xy}^2)$ and the temperature dependence is shown in Fig.~\ref{fig:transport}(b). 

The anomalous Hall conductivity saturates to $\sigma_{xy}\simeq500~\Omega^{-1}$cm$^{-1}$ at 2~K, consistent with previous studies \cite{Meng2019, Yang2020, Destraz2020, Sanchez2020}. This value is close to the expected value from intrinsic Berry curvature contribution $\sigma_{xy}=\frac{2\pi e^2}{\hbar a}\simeq 600~\Omega^{-1}$cm$^{-1}$, where $\hbar$ is the reduced Planck's constant, $e$ is the electric charge and $a$ is the lattice constant. {\color{black} While a conventional FM order is expected to give rise to a finite $\sigma_{xy}$ due to extrinsic effects such as skew scattering or sider jump, the amplitude is around $0.1 - 0.001$ times of $\frac{2\pi e^2}{\hbar a}$ ($\sim 0.6 - 60~\Omega^{-1}$cm$^{-1}$) in PrAlGe \cite{Meng2019}, which is much smaller than observed (Fig.~\ref{fig:transport}(b)).} Most importantly however, {\color{black}large} finite values of $\sigma_{xy}$ is observed above $T_C$, up to 30~K. These kind of observations has in the past been used to assert TRSB and AHE in different kinds of systems. We shall however show below that the observed behaviour in this case can be ascribed to a magnetic field induced effect. 

The large intrinsic anomalous Hall conductivity is explained within the framework of Berry curvature and finite Chern number. A prerequisite is a nonzero sum of Berry curvature, which in turn relies on TRSB. In PrAlGe, the inversion symmetry is broken by the lattice symmetry and the time reversal symmetry is broken below $T_{\rm C} = 15$~K, which is expected to shift the Weyl points in $k$-space. This leads to a finite $k$ range where sum of Berry curvature becomes nonzero (due to the Zeeman coupling between the localized $f$-electrons and the Weyl points), thereby realizing large intrinsic anomalous Hall conductivity in the FM phase (see supplementary materials {\color{blue}\cite{SM}}) \cite{Chang2018, Deng2020}. 

Above $T_{\rm C}$ on the other hand, the TRS is preserved as confirmed by $\mu^+$SR. Yet, zero field extrapolated Hall conductivity is unequivocally measured (Fig.~\ref{fig:transport}(b)). We can explain this behaviour theoretically by combining a simple model for a Weyl semimetal subject to a Zeeman field $\boldsymbol{B}(\boldsymbol{H},T)=\mu_0(\boldsymbol{M}(\boldsymbol{H}, T)+\boldsymbol{H})$. For $\boldsymbol{M}(\boldsymbol{H}, T)$ we take the behaviour that follows from  mean-field theory of a ferromagnet (see Fig.~\ref{fig:transport}(d)). Assuming that the main temperature dependence of the anomalous Hall conductivity stems from the temperature dependence of $\boldsymbol{M}(\boldsymbol{H}, T)$, we obtain the Hall conductivity by integration over the Berry curvature of occupied states. The result, shown in Fig.~\ref{fig:transport}(c) reproduces the qualitative behaviour of the experimental data above, at, and below $T_\mathrm{C}.$

In summary, we studied the intrinsic anomalous Hall effect (AHE) using the ferromagnetic Weyl semimetal PrAlGe as a case study. This AHE was derived from zero-field extrapolation of the field-dependent Hall resistance. Consistent with previous results, finite values are observed below $T_{\rm C} = 15$ K. Additionally, we present clear evidence that a similar extrapolated response is also found in the paramagnetic phase, challenging the conventional understanding that such a response is synonymous with AHE and thereby a definitive indicator of time-reversal symmetry breaking (TRSB). Muon spin relaxation measurements confirmed a paramagnetic phase above $T_{\rm C}$ without TRSB. Instead, by employing a mean-field theory approach, we explain the observed temperature dependence as a result from induced magnetisation due to the applied magnetic field {\color{black} in presence of strong ferromagnetic fluctuations}. Our findings have profound implications for the study of AHE in systems where TRSB is contentious, such as Kagome metals, potentially reshaping the discourse in the field. 
\\~\\
\textbf{\label{sec:conclusions}Acknowledgments}
O.K.F. is supported by the Swedish Research Council (VR) via a Grant 2022-06217 and the Foundation Blanceflor 2023 and 2024 fellow scholarships. X.L. acknowledges support from the National Key R$\And$D Program of China (Grant No. 2022YFA1403700). M.H., K.K., Q.W. and J.C. acknowledge support from the Swiss National Science foundation (SNSF) under grant numbers 200021$\_$188564 and BSSGI0$\_$155873. Q.W. is supported by the Research Grants Council of Hong Kong (ECS No. 24306223). We acknowledge Diamond Light Source for time on Beamline I05 under Proposal SI22091. S.S., T.S. and E.P. acknowledge support from the SNSF under grant No. 200021-188706. J.W. and V.U. acknowledge support from the SNSF Projects 200021-188707 and Sinergia CRSII5-171003 NanoSkyrmionics. T.S. also acknowledge the Natural Science Foundation of Shanghai (Grants No.\ 21ZR1420500 and 21JC\-140\-2300) and National Natural Science Foundation of China (Grant No. 12374105)

\normalem
\bibliography{Refs} 

\clearpage 


\onecolumngrid 
\begin{center}
    \textbf{End Matter} 
\end{center}
\twocolumngrid 

{\color{black}
 \textit{Appendix A: Electrical transport} -- A rectangular shaped flux-grown PrAlGe single crystal was prepared for electrical transport. Four Au-wires were attached to the ab-plane by silver paste to apply current (1 mA) and magnetic field along the crystallographic $a$- and $c$-axis, respectively, confirmed by x-ray Laue. Electrical resistivity ($\rho_{xx}$) and Hall resistivity ($\rho_{xy}$) were measured with both positive and negative fields to eliminate effects of contact misalignment. Figure~\ref{fig:p_xy}(a,b) show magnetic field dependence of $\rho_{xy}$ for selected temperatures. The intercept at zero field of the linear behaviour extrapolated from higher fields defines $\rho^A_{xy}$ and thereby $\sigma_{xy} = -\rho^A_{xy}/(\rho_{xx}^2+\rho_{xy}^2)$. 

\begin{figure}[hb!]
  \begin{center}
     \includegraphics[keepaspectratio=true,width = 90 mm]{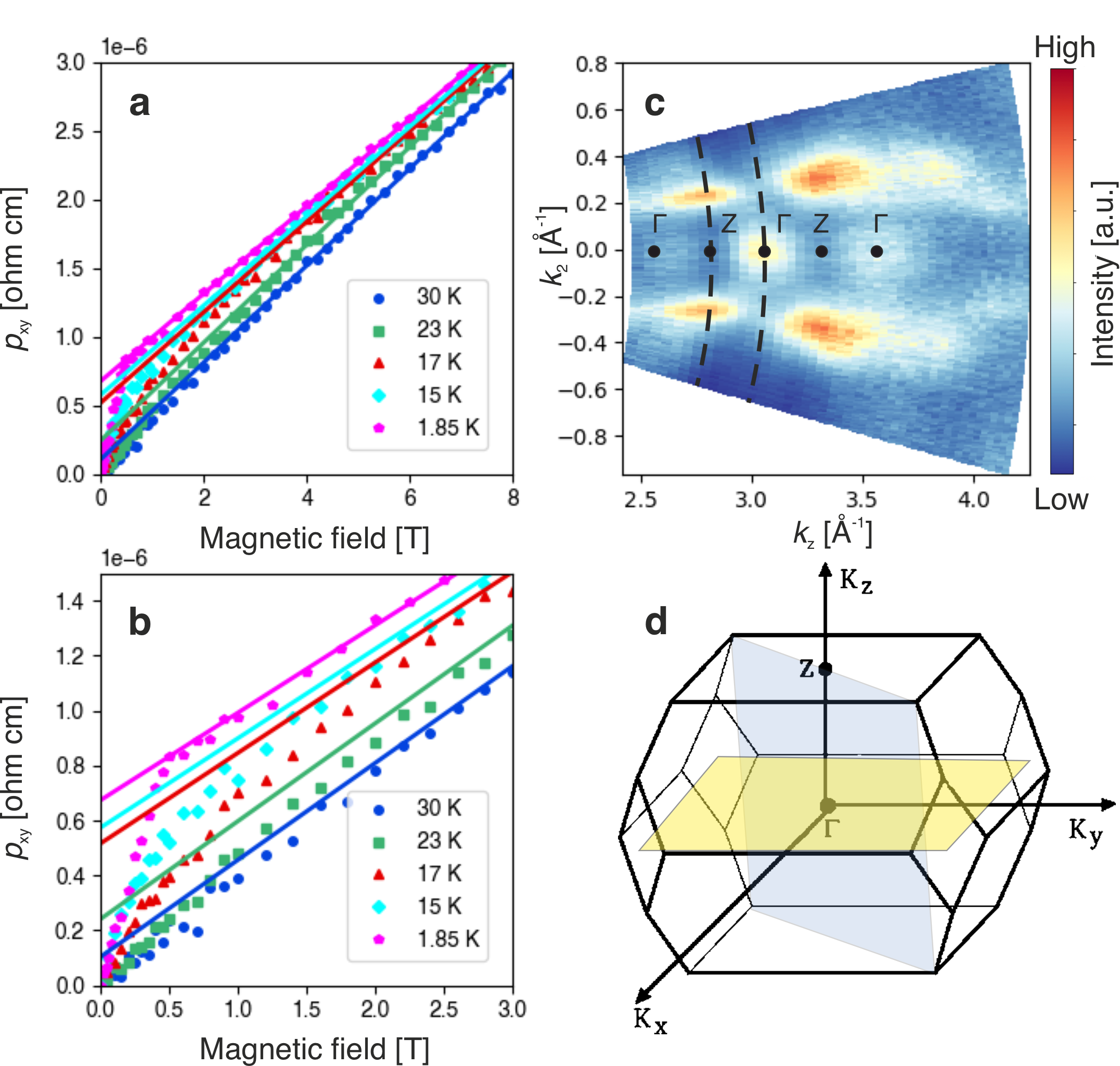}
 \end{center}
    \caption{{\color{black}(a, b) Magnetic field dependence of the Hall resistivity ($\rho_{xy}$) for selected temperatures. The solid lines are linear fits between 2.5 to 8~T, extrapolated to zero field. (c) Out of plane Fermi surface along the (001) crystal axis, measured along in-plane $\Gamma$-M direction. We have defined $k_2 = k_x + k_y$, according to the main text. The dashed lines represents the Brillouin zone measured in this study. (d) Brillouin zone of PrAlGe \cite{Chang2018}. Blue shaded area highlights the scan direction of {\color{blue}(c)}. The yellow shaded area represents the Fermi surface presented in Fig.~\ref{fig:ARPES}(a,b).}
        }
    \label{fig:p_xy}
\end{figure}

\begin{table*}[ht]
\caption{\label{table:table}
{\color{black}Summary of samples used in each experiment. For reference, previous experimental studies cited in this work, which are based on different growth methods, are listed as well. }
}
\begin{ruledtabular}
\begin{tabular}{lccccccc}
 Growth method & Crystal form & Experiments & $T_{\rm C}\simeq$ & References\\
 
  \hline
 Al-Flux &  Single crystal & Hall effect, Magnetic susceptibility & 15~K &\cite{Puphal2019, Meng2019, Destraz2020, Sanchez2020} \\
 
 Floating zone & Single crystal & ARPES & 15~K  &\cite{Puphal2019, Destraz2020}\\
  
 Arc melting  & Polycrystal & $\mu^+$SR & 15~K &\cite{Puphal2019, Puphal2020} \\
 
\end{tabular}
\end{ruledtabular}
\end{table*}

\vspace{1em}

\textit{Appendix B: Magnetic susceptibility} -- The temperature-dependent magnetic susceptibility of the flux-grown single-crystal PrAlGe was measured on the magnetic property measurement system (MPMS, Quantum Design) equipped with the superconducting quantum interference device. The crystal was zero field cooled to base temperature, at which an excitation field of 1000~Oe was applied. 

\vspace{1em}

\textit{Appendix C: Angle resolved photo emission spectroscopy} -- Experiments were performed at the I05 beamline at DIAMOND light source above and below the FM ordering temperature $T_{\rm C} = 15$~K. Floating-zone grown PrAlGe crystals were cleaved in-situ and measured in ulta-high vacuum of $\sim 2\times10^{-10}$~mbar with a $50 ~\mu$m beam spot. All presented spectra were collected with linear horizontal polarisation on the (001) surface. Fermi level and spectral ARPES intensities were calibrated based on measurements from electrically contacted polycrystalline gold. To orient within the Brillion zone, a photon energy ($h\nu$) scan was performed from 20 to 64~eV at $T = 10$~K. The out of plane Fermi surface along the (001) crystal axis, measured along in-plane $\Gamma$-M direction is shown in Fig.~\ref{fig:p_xy}(c). The photon energy was converted into $k$-space via the relationship $k_z = 
 \frac{\sqrt{2m_eE_{k}(cos^2\theta+V_0)}}{\hbar}$, where $m_e$ is the mass of the electron, $E_k$ is the kinetic energy of emitted electron, $\theta$ is the emission angle, $\hbar$ is the reduced Planck constant and $V_0 = 10$~eV is the inner potential. Each spectra have been normalised based on the total accumulated intensity. Fig.~\ref{fig:p_xy}(d) shows the Brillouin zone of PrAlGe~\cite{Chang2018} which highlights the relevant directions in this study.

\vspace{1em}

\textit{Appendix D: Muon spin relaxation} -- The $\mu^+$SR measurements were conducted on the GPS instrument at Paul Scherrer Institute (PSI). Muon spin relaxation time spectra were collected in ZF cooled configuration using a low background Cu fork holder inserted into a flow He cryostat reaching down to 1.6~K. Powderized polycrystalline ingots of PrAlGe were pressed into pellets of 5~mm in diameter. $\mu^+$SR relies on implanting nearly 100\% spin-polarised muons, acting as a local magnetometer. The high gyromagnetic ratio of the muon allows for measurements of weak (a few~Oe) magnetic fields, and are sensitive to both static and dyanmic magnetic field in the range of MHz-GHz. The muon data was analysed using $musrfit$ \cite{musrfit}, and descriptions of the technique are found elsewhere \cite{Yaouanc2011, Forslund2021_PHD}. 

Muon spin relaxation ($\mu^+$SR) time spectra were collected as a function of temperature and longitudinal field {\color{blue}(LF)}. The measured zero field (ZF) time spectra for selected temperatures is shown in Fig.~\ref{fig:ZF}(a). An exponential like relaxation is presented for all $T>T_{\rm C}=15$~K. Therefore, the time spectra were fitted using an exponentially relaxing longitudinal field static Gaussian Kubo-Toyabe {\color{black}(LFSGKT)} function

\begin{eqnarray}
 A_0 \, P_{\rm ZFLF}(t)= 
 AG^{\rm LFSGKT}(\Delta,t, B_{\rm LF})e^{-\lambda t}
\label{eq:ZF}
\end{eqnarray}

where $A_0$ is the initial asymmetry, a value determined by the instrument, and $P_{\rm ZFLF}$ is the muon spin relaxation function under LF and ZF (with $B_{\rm LF}=0$~Gauss). $G^{LFSGKT}$ represent the longitudinal field static Gaussian Kubo-Toyabe function, where $\Delta$ is the field distributing width and $\lambda$ is the exponential relaxation rate. {\color{black}Setting $B_{\rm LF}=0$~G reduces $G^{LFSGKT}$ into static Gaussian Kubo-Toyabe ($G^{SGKT}$) function}. The response of a muon ensemble originating from two independent magnetic field distributions is given by the Fourier transform of their convolution. Therefore, the total response can be modeled as a product of each response. In this case, the exponential relaxation entails fluctuating electronic moments whereas $G^{SGKT}$ represents field distributions arising from mostly $I_{\rm Pr}=5/2$ and $I_{\rm Al}=5/2$ randomly oriented nuclear moments. 

Figure~\ref{fig:ZF}(c) shows $\lambda$ and $\Delta$ as a function of temperature obtained in ZF through two different fitting procedures. $\lambda_{\rm free}$ and $\Delta_{\rm free}$ were obtained using a fit procedure with no restrictions made. In this case, $\Delta_{\rm free}$ increases as $\lambda_{\rm free}$ increases at lower temperatures. This kind of behaviour is however nonphysical, assuming that there is no coupling between nuclear and electronic moments. Instead, this increase is a fitting artifact where the value of $\Delta_{\rm free}$ is obscured by $\lambda_{\rm free}$, as $\lambda_{\rm free}$ increases to very high values ($> 1~\mu$s$^{-1}$). In fact, a single exponential can fit the ZF spectra at lower temperatures $< 20$~K and the KT is not needed to reproduce the spectra. Although, a nuclear depolarisation is in principle always there and including the KT relaxation represents a more correct physical description.  

To account for the described fitting defect, a proper value of $\Delta$ needs to be estimated. The relaxation rate ($\lambda$) decreases with increasing temperature. Therefore, all ZF time spectra collected at temperatures above 20~K were simultaneously fitted with having $\Delta$ as a shared parameter, in which $\Delta\simeq~$15~$\mu$s$^{-1}$ was obtained. The presented $\lambda$ in Fig.~\ref{fig:muon}(d) is the obtained temperature dependence with $\Delta=~15~\mu$s$^{-1}$ fixed. It is noted that regardless of the fitting procedure, very similar temperature dependencies are obtained for the exponential relaxation rate (Fig.~\ref{fig:ZF}(c)).  

Figure~\ref{fig:ZF}(b,d) shows measured muon spin time spectra and the corresponding relaxation rate ($\lambda$) as a function of {\color{blue}LF} collected at $T = 20$~K. All time spectra were fitted simultaneously by having $\Delta = 0.15$~$\mu$s$^{-1}$ and the asymmetry ($A$) as a shared parameter, effectively reducing the number of fit parameters to one ($\lambda$). Apart from the initial decoupling (the decrease in $\lambda$), there is no significant LF dependence observed. {\color{blue} The initial decrease in relaxation rate can be mostly attributed to nuclear moments, though a small fraction of static or slowly fluctuating fields may be present as well. However, no further decrease is observed beyond the initial decoupling, indicating that the dominant fluctuations persist up to at least 5000~G.}

Below $T_{\rm C}$, the time spectra exhibits a rapid relaxation followed by a slow relaxation (Fig.~\ref{fig:muon}(c)). In a perfect powder, 2/3 of the internal magnetic field is expected to be perpendicular and 1/3 of the internal magnetic field is expected to be parallel with respect to the initial muon spin polarisation. However, the time spectra below $T_{\rm C}$ do not exhibit this ratio between the parallel and perpendicular components. This is because the polycrystals used in this study exhibits a preferred orientation. The tetragonal crystals have a strong $c$-axis preference, which is the FM order direction. This causes a higher asymmetry (signal) in the parallel component compared to the perpendicular one.

\begin{figure}[ht]
  \begin{center}
     \includegraphics[keepaspectratio=true,width=90 mm]{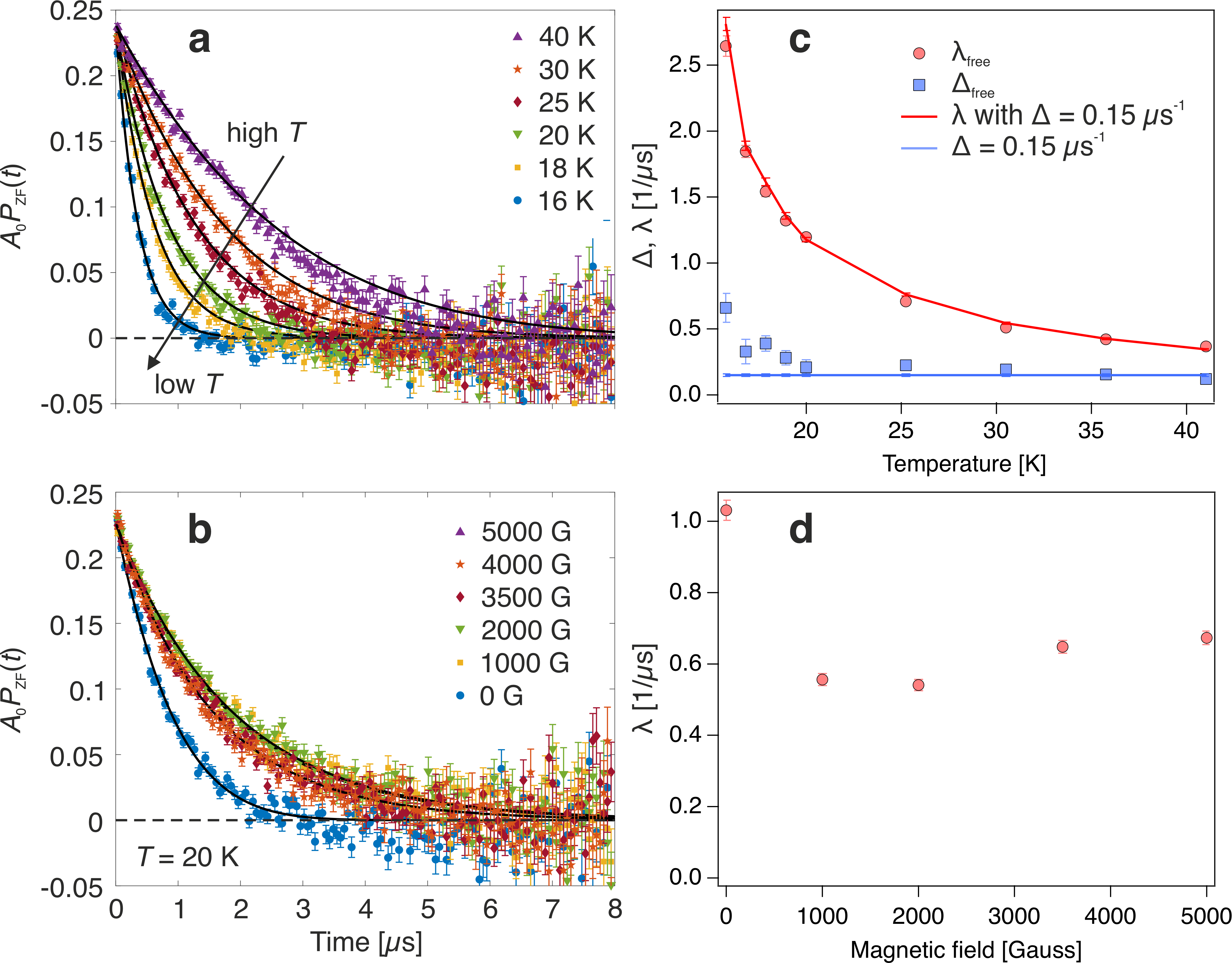}
 \end{center}
    \caption{{\color{black} (a, b) Zero field and longitudinal field muon spin relaxation time spectra. The solid lines represent best fits using Eq.~\ref{eq:ZF} with fixed $\Delta=0.15~\mu$s$^{-1}$. (c) $\lambda$ and $\Delta$ as a function of temperature obtained through two different fitting procedures. The scattered points ($\lambda_{\rm free}$ and $\Delta_{\rm free}$) were obtained using a fit procedure with no restrictions made. The solid lines were obtained by fixing $\Delta=~0.15~\mu$$s^{-1}$. (d) $\lambda$ as a function of longitudinal fields, measured at $T = 20$~K. }
        }
    \label{fig:ZF}
\end{figure}

\end{document}